\documentclass[3p,times,twocolumn]{elsarticle}

\usepackage{ecrc}

\volume{00}

\firstpage{1}

\journalname{Nuclear and Particle Physics Proceedings}

\runauth{}

\jid{nppp}

\jnltitlelogo{Nuclear and Particle Physics Proceedings}

\usepackage{multicol}

\usepackage{amssymb}
\usepackage[figuresright]{rotating}

\usepackage{float}
\begin{document}

\begin{frontmatter}



\dochead{}

\title{ Pre-equilibrium Longitudinal Flow in the IP-Glasma Framework for Pb+Pb Collisions at the LHC
}


\author[McGill]{Scott McDonald}

\author[McGill,Brookhaven]{Chun Shen}
 
\author[INRS,Waterloo]{Fran\c{c}ois Fillion-Gourdeau}

\author[McGill]{Sangyong Jeon}
 
 \author[McGill]{Charles Gale}
 
 \address[McGill]{Department of Physics, McGill University, 3600 University
 Street, Montreal,
 QC, H3A 2T8, Canada}
 
 \address[Brookhaven]{Physics Department, Brookhaven National Laboratory, Upton, NY 11973, USA}
 
\address[INRS]{Universit\'e du Qu\'ebec, INRS-\'Energie, Mat\'eriaux et
T\'el\'ecommunications, Varennes, Qu\'ebec, Canada J3X 1S2}

\address[Waterloo]{Institute for Quantum Computing, University of Waterloo, Waterloo, Ontario, Canada, N2L 3G1}

\begin{abstract}
In this work, we debut a new implementation of IP-Glasma and quantify the pre-equilibrium longitudinal flow in the IP-Glasma framework. The saturation physics based IP-Glasma model naturally provides a non-zero initial longitudinal flow through its pre-equilibrium Yang-Mills evolution. A hybrid IP-Glasma+MUSIC+UrQMD framework is employed to test this new implementation against experimental data and to make further predictions about hadronic flow observables in Pb+Pb collisions at 5.02 TeV. Finally, the non-zero pre-equilibrium longitudinal flow of the IP-Glasma model is quantified, and its origin is briefly discussed.  
 
\end{abstract}


\end{frontmatter}

% \linenumbers


\section{Introduction}
Quark Gluon Plasma (QGP), a near perfect fluid of deconfined quarks and gluons, is believed to be formed in heavy ion collisions conducted at RHIC and the LHC. Phenomenological studies seek to constrain the parameters of the existing theoretical models by comparison to experimental data, and Run 2 at the LHC provides a new opportunity to do so. In this work, the IP-Glasma+MUSIC+UrQMD hybrid model will be employed to study Pb+Pb collisions at 2.76 TeV and 5.02 TeV. In creating a new implementation of IP-Glasma \cite{1202.6646, 1206.6805}, without altering the underlying physics, we aim to further explore the parameter space of the model. Before testing the efficacy of this implementation it is useful to briefly review some key features of IP-Glasma. 

\section{IP-Glasma Implementation}
\label{}

The IP-Glasma model determines the saturation scale for the small-x gluons via the IP-Sat model \cite{hep-ph/0304189}, which is related to the color charge density through $Q_s\approx 0.5g^2\mu$.
%%%%%%%%%%%%%%%%
%\begin{equation}
%\end{equation}
%%%%%%%%%%%%%%%%
Color charge fluctuations inside each nucleus are then sampled at each position in the transverse plane using the MV model \cite{hep-ph/9311205}
\noindent\begin{equation}
\langle \rho_{A(B)}^{a}(\mathbf{x_{\perp}})\rho_{A(B)}^{b}(\mathbf{y_{\perp}}) \rangle=g^{2}\mu_{A(B)}^{2}(x,\mathbf{x_{\perp}})\delta^{ab}\delta^{2}(\mathbf{x_{\perp}}-\mathbf{y_{\perp}})
\end{equation}
The gauge fields of the two incoming nuclei, denoted as A and B, respectively, are obtained by solving the Yang-Mills equations in the Lorentz gauge, which reduce to the 2-dimensional Poisson equation.

The gauge fields are initialized on the light cone with the usual Color Glass Condensate \cite{hep-ph/9502289} boundary conditions,

\begin{minipage}{.5\linewidth}
\begin{equation}
  A^{\eta}=\frac{ig}{2}[A^{i}_{(A)},A^{i}_{(B)}]
\end{equation}
\end{minipage}%
\begin{minipage}{.5\linewidth}
\begin{equation}
 A^{i} = A^{i}_{(A)} + A^{i}_{(B)}
\end{equation}
\end{minipage}
and evolved in the forward light-cone using the sourceless classical Yang-Mills equations,
\begin{equation}
    [D_{\mu}, F^{\mu\nu}]=0
\end{equation}
as implemented on the lattice \cite{hep-ph/0303076}. After evolving to a switching time of $\tau_{sw} =0.4 \, {\rm fm}$, the stress energy tensor is constructed from the chromo-magnetic and chromo-electric fields \cite{hep-ph/0605045}, 
\begin{equation}\label{tmunu}
  T^{\mu\nu} = -g^{\mu\alpha}g^{\nu\beta}g^{\gamma\delta} F_{\alpha\gamma}F_{\beta\delta} + 
  \frac{1}{4}g^{\mu\nu}g^{\alpha\gamma}g^{\beta\delta} F_{\alpha\beta}F_{\gamma\delta}
\end{equation}
where $g^{\mu\nu} = {\rm diag}(1, -1, -1, -1/\tau^2)$ in $\tau$-$\eta$ coordinates.
Equation (\ref{tmunu}) is diagonalized
\begin{equation}\label{eigen}
    T^{\mu}_{\,\, \nu} u^{\nu} = \epsilon u^{\mu}
\end{equation}
to yield the hydrodynamic quantities $u^{\mu}$ and $\epsilon$. These represent the local flow velocity and energy density, respectively.

\section{Centrality Determination}\label{centrality}
The IP-Glasma model is made up of small-$x$ gluons, that can be approximated as classical gluon fields due to their high occupation number. These gluon fields can interact with one another at finite separation.  Such interactions leave some freedom in the criteria for whether a collision took place. In this work, we relied on experimental data to fix this ambiguity.  

\begin{figure}
    \centering
    \includegraphics[width=0.8\linewidth]{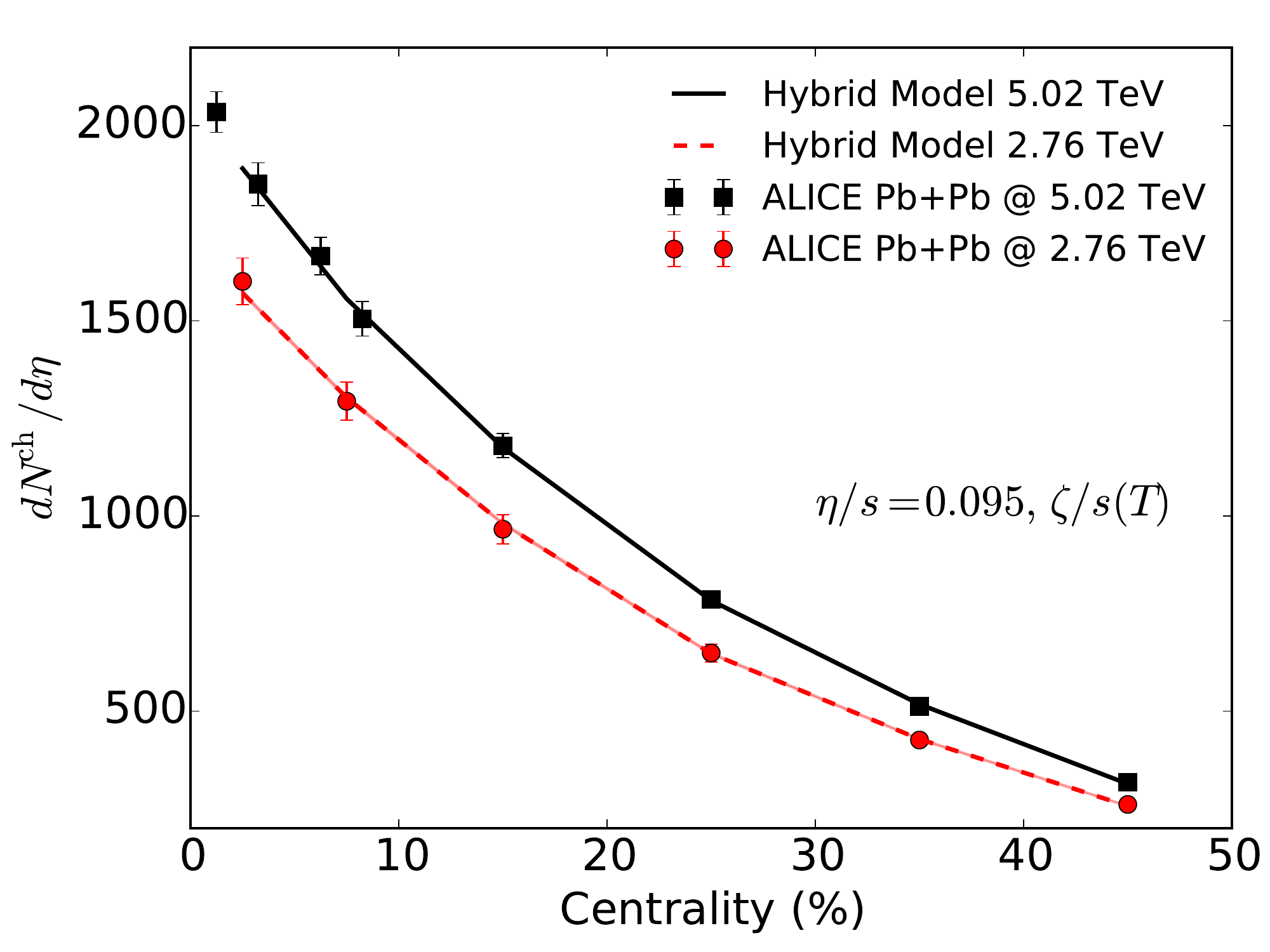}
    \caption{Charged hadron $dN^{\rm{ch}}/d\eta$ from the procedure described in section \ref{centrality} from our IP-Glasma+MUSIC+UrQMD hydrid model compared to ALICE data \cite{Aamodt:2010cz, Adam:2015ptt}.}
    \label{fig:dN}
\end{figure}

The centrality selection procedure, described in \cite{1609.02958}, begins by sampling on the order of $\approx$ 25K IP-Glasma events per collision energy in the impact parameter range $0 \leq b \leq 20$. Then MUSIC, a second-order relativistic  viscous  hydrodynamics code \cite{1009.3244}, was run using a subset of these initial conditions.  From these events, the initial state energy density $\frac{dE}{\tau d\eta_s}\vert_{\tau=0.4}$, in GeV/fm, was related to the final state charged hadron multiplicity $\frac{dN_\mathrm{ch}}{d\eta}\vert_{\vert \eta \vert < 0.5}$ using a power law fit, where 
$\eta$ is pseudo-rapidity and $\eta_s$ is space-time rapidity. Ultimately, the resulting relationship,
\begin{equation} \label{eq:dNdy}
\frac{dN_\mathrm{ch}}{d\eta}\bigg\vert_{\vert \eta \vert < 0.5}=0.839\left( \frac{dE}{\tau d\eta_s}\bigg\vert_{\tau=0.4 \, {\rm fm}}  \right)^{0.833}
\end{equation}
was used to bin the IP-Glasma events such that they reproduced the experimental ratios of $dN^{ch}/d\eta$  between neighboring centrality bins. The result of this procedure can be seen in Fig. (\ref{fig:dN}).
%%%%%%%%%%%
The study consisted of 1500 events per 10\% centrality class at each collision energy. 

\section{Model Setup}
From the quantities obtained from equation (\ref{eigen}), Landau matching is used to initialize hydrodynamic simulation by reconstructing the ideal hydrodynamic stress-energy tensor from the quantities obtained via equation (\ref{eigen}). The parameters for MUSIC are based on the parametrization used in \cite{1502.01675}. By choosing this parametrization, which has proven capable of describing a wide range of observables, we were able to both isolate the effects of this new implementation of IP-Glasma, and to avoid tinkering with parameters. To summarize, we employed the s95p-v1 equation of state \cite{0912.2541}, a constant shear viscosity of $\eta/s = 0.095$, and a temperature dependent bulk viscosity. The temperature dependent bulk viscosity parametrization that was developed in \cite{0903.3595}  was utilized with a slight modification. The peak was reduced to 90\% of its original value to account for a slight difference in initial flow from our IP-Glasma implementation. 

As the fluid expands and cools, the system undergoes hadronization, at which point we switch to UrQMD \cite{nucl-th/9803035} to handle the hadronic cascade.  The switching temperature was taken to be $T_{sw}=145 \, {\rm MeV}$ and the default UrQMD parameters were used.
%%%%%%%%%%
\begin{figure}[b]
  \centering 
    \includegraphics[width=0.8\columnwidth]{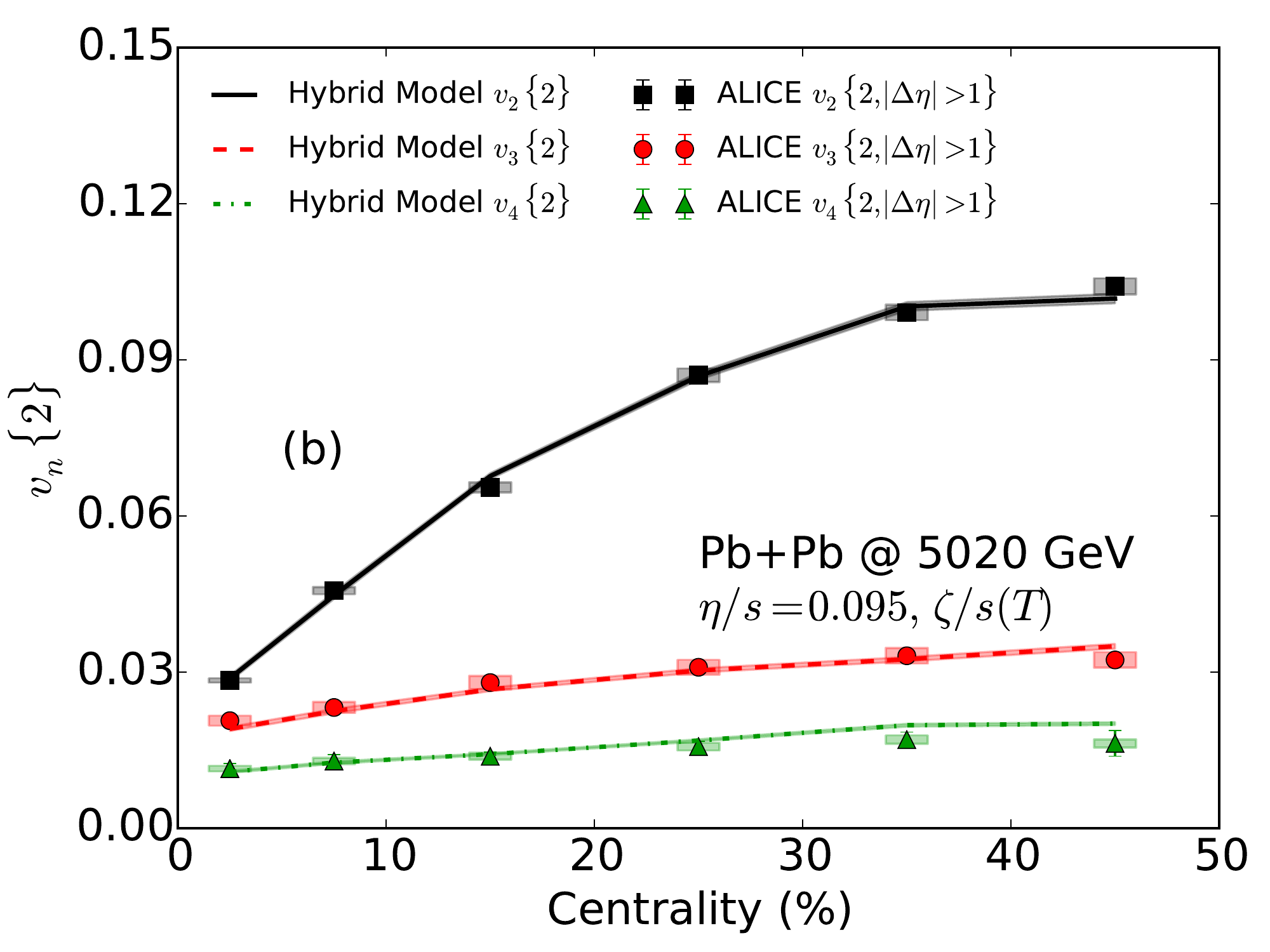}
    \caption{Integrated charged hadron $v_n$ at 5.02 TeV compared to ALICE data \cite{1602.01119}.}
    \label{fig:5.02}
\end{figure}

Importantly, the same MUSIC and UrQMD parameters have been used  for both energies presented in \cite{1609.02958}. The ability of this hybrid model to describe data at both $2.76 \, {\rm TeV}$ and $5.02 \, {\rm TeV}$  suggests that similar temperatures are probed at the two collision energies.  To make a more forceful statement, it would be necessary to do a thorough study of the temperature evolution of the medium at both energies.  
\section{Hadronic Flow Observables at LHC Energies}
To test this new implementation, we consider a number of well-known results from IP-Glasma. As a baseline, the charged hadron integrated $v_n$ are computed, as shown for 5.02 TeV in Fig. (\ref{fig:5.02}) and good agreement with data is found. The two energies considered in \cite{1609.02958} had very similar initial state energy anisotropies, $\epsilon_n$'s, but the events at 5.02 TeV had consistently higher ratios of $\langle v_n\rangle /\langle \epsilon_n\rangle$ across centralities. Thus the increased $v_n$ values that are found at the higher energy are largely attributable to the increased lifetime of the fireball.

\begin{figure}[h!]
    \centering    
    \includegraphics[width=0.8\linewidth]{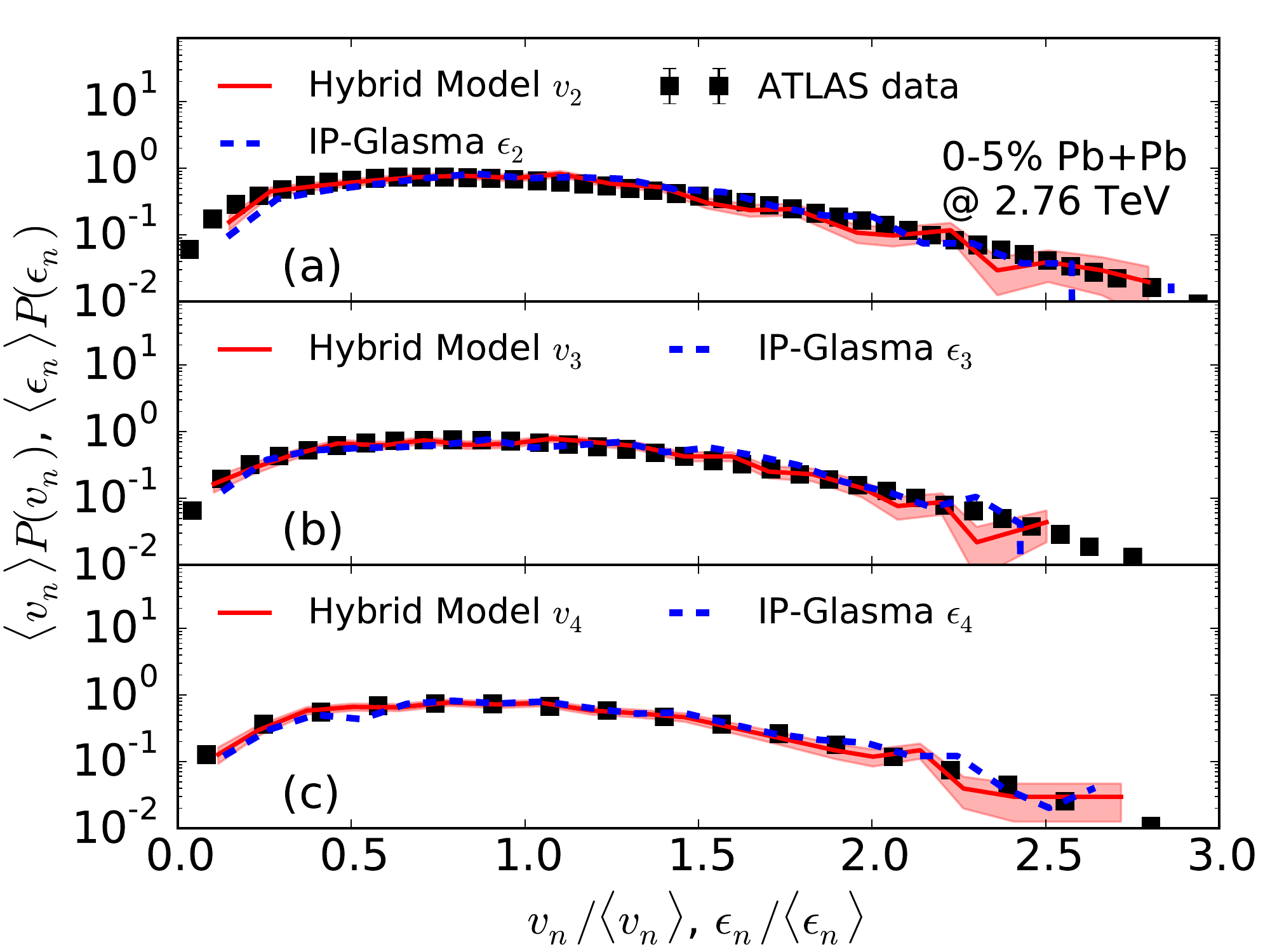}
    \caption{The normalized distributions of initial state energy anisotropies, $\epsilon_n$, the anisotropic flow coefficients $v_n$, as compared to the experimental normalized distribution of $v_n$ from ATLAS \cite{1305.2942} for the 0-5\% centrality bin. }
    \label{vndis}
\end{figure}

Beyond event-averaged $v_n$ harmonics, IP-Glasma's event-by-event fluctuations are able to reproduce the experimental $v_n$ distributions \cite{1209.6330}. This finding is reproduced in Fig. (\ref{vndis}).
%%%%%%%%%%%
In \cite{1609.02958}, many more observables from this study are computed and compared with data, and good agreement between the theoretical and experimental results is achieved. We take this as evidence of the efficacy of our new IP-Glasma implementation and the model generally.
\section{Non-zero Pre-equilibrium Longitudinal Flow}
\begin{figure}[h!]
    \centering
    \includegraphics[width=0.9\linewidth]{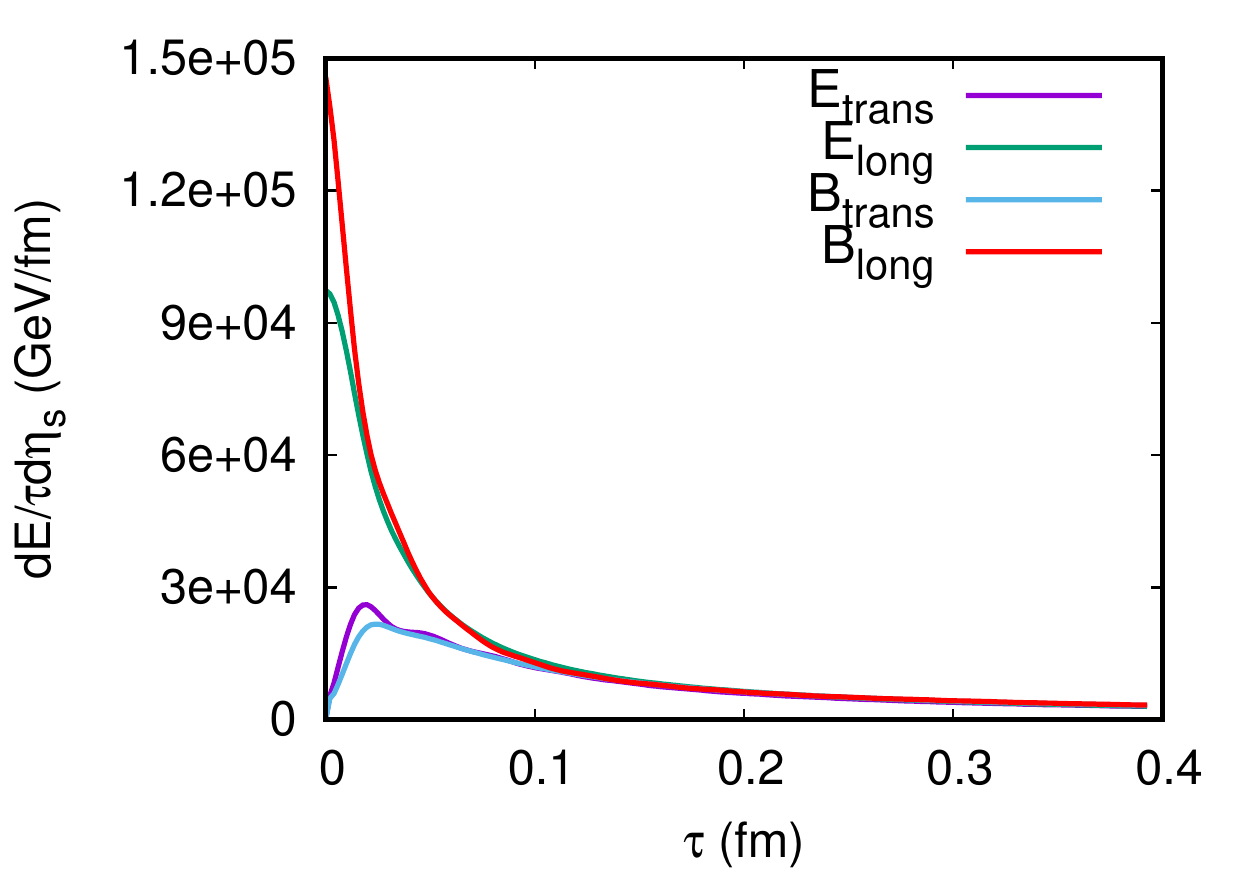}
    \caption{Energy density contributions from the different components of the chromo-magnetic and chromo-electric fields as a function of time.}
    \label{fig:energy_density}
\end{figure}

\begin{figure}[h!]
    \centering
    \includegraphics[width=1.0\linewidth]{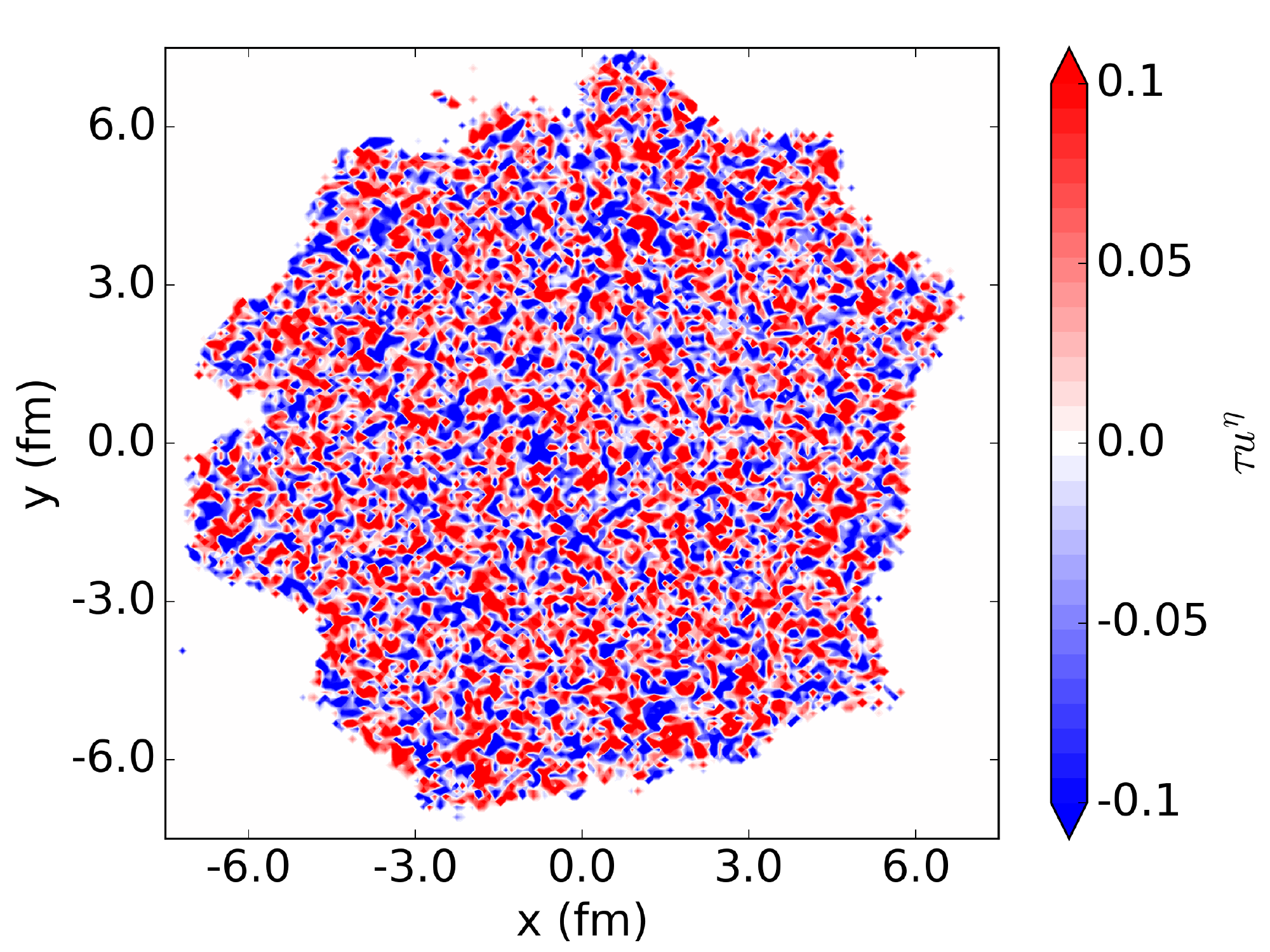}
    \caption{Pre-equilibrium longitudinal flow from an IP-Glasma event in the $0\textnormal{-}5\%$ centrality bin at $\tau=0.4 {\rm \, fm}$.}
    \label{fig:long_flow}
\end{figure}

Owing largely to the historical fact that 2+1D simulations have preceded their 3+1D counterparts, hydrodynamic simulations have almost always been initialized with vanishing initial longitudinal flow, $u^{\eta}=0$. 
%%%%%%%%%%%%%%%%%%%%%%%%
\begin{figure}[t]
    \centering
    \includegraphics[width=1.\linewidth]{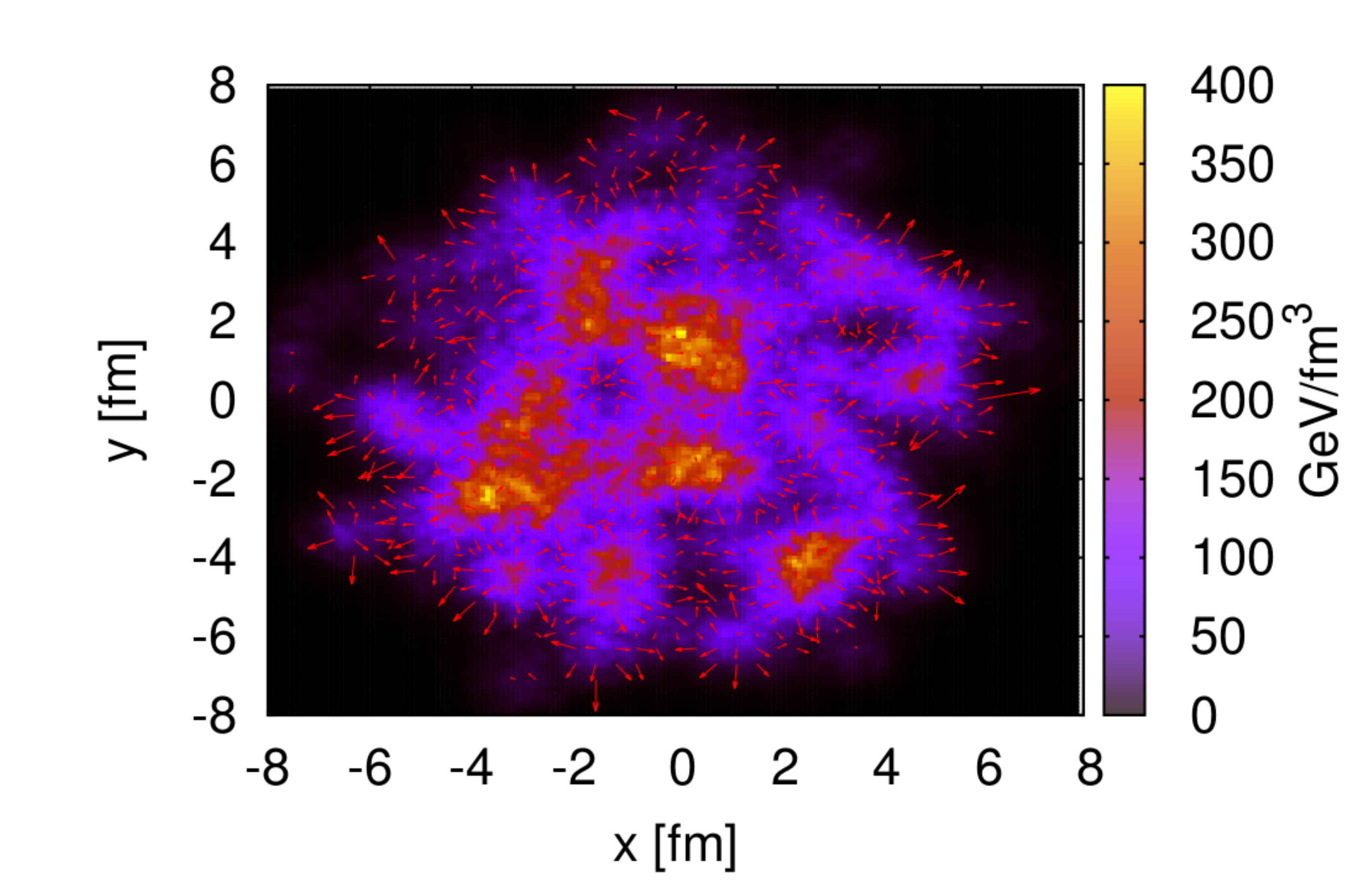}
    \caption{The transverse flow field superimposed on the energy density from an IP-Glasma event in the $0\textnormal{-}5\%$ centrality bin at $\tau=0.4 \, {\rm fm}$.}
    \label{fig:transverse_flow}
\end{figure}
Thus our conception of pre-equilibrium flow has been confined to its transverse components, such as those plotted in Fig. (\ref{fig:transverse_flow}).  However, even in the boost invariant case of IP-Glasma, the form of the chromo-electric and chromo-magnetic fields lead to a stress energy tensor that, when diagonalized, will yield a non-zero $u^{\eta}$.  Equation (\ref{tmunu}) will have components in the $\eta$-direction such as,
\begin{equation}
    T^{\tau\eta} = F^{\tau x}F^{\eta x} + F^{\tau y} F^{\eta y} \neq 0 
\end{equation}
Consider Fig. (\ref{fig:energy_density}), for example.
At typical switching times, the energy contribution from the longitudinal and transverse fields are of the same order, and thus the $\eta$-components of $T^{\mu\nu}$ are non-zero. Thus, solving the eigenvalue problem yields a non-zero $u^{\eta}$, as seen in Fig. (\ref{fig:long_flow}).  In fact, by considering $\tau u ^{\eta}$, in order to have quantities with the same units, one can compare the longitudinal and transverse flow. 
Looking at 100 events at $2.76$ TeV in the $0\textnormal{-}5 \%$ centrality bin, and using an RMS definition to quantify each component of the flow field, 
%%%%%%%%%%%%%
\begin{equation}
\langle u^{\mu} \rangle= \sqrt{\frac{\int{(u^{\mu})^2\epsilon d^2x}}{\int{\epsilon d^2x}}},
\end{equation}
%%%%%%%%%%%%%
the longitudinal flow is found to be of the same order of magnitude as the transverse flow, $\langle \tau u^{\eta} \rangle \approx 0.5\langle u^{\perp}\rangle$.
%%%%%%%%%%%
The Landau matching procedure that is normally used to reconstruct the ideal hydrodynamic stress-energy tensor from the local flow field $u^{\mu}$ and energy density $\epsilon$ neglects the viscous shear tensor $\pi^{\mu \nu}$. In the absence of isotropization this procedure has become standard, but certainly not ideal. By neglecting the initial $u^{\eta}$ we are throwing away yet more information from the Yang-Mills evolution. Future phenomenological studies should include this component, particularly as the community moves towards 3+1D simulations. 
 
We expect that the effects of non-zero initial longitudinal flow will likely have the largest effect on photons since thermal photon yields receive an important contribution from the early QGP evolution, when temperatures are highest. This represents an opportunity to test the phenomenological effects of including this component of the flow velocity.
\section {Conclusion}
We have developed a new numerical implementation of IP-Glasma and demonstrated its efficacy by comparing with experimental data for Pb-Pb collisions at 2.76 TeV and 5.02 TeV. The non-zero pre-equilibrium longitudinal flow generated in the IP-Glasma framework was quantified in relation to the transverse flow. When matching models with pre-equilibrium flow to hydrodynamics, it is best to retain the initial longitudinal flow if at all possible, in order to avoid discarding information from the pre-equilibrium dynamics. Plans for future work include studying the phenomenological effects of non-zero initial longitudinal flow in hydrodynamical evolutions.
\section{Acknowledgements}
This work was supported in part by the Natural Sciences and Engineering Research Council of Canada, as well as the U.S. Department of Energy, Office of Science under contract No. DE- SC0012704. Computations were made in part on the supercomputer Guillimin from McGill University, managed by Calcul Qu\'ebec and Compute Canada. The operation of this supercomputer is funded by the Canada Foundation for Innovation (CFI), NanoQu\'ebec, RMGA and the Fonds de recherche du Qu\'ebec - Nature et technologies (FRQ-NT). C. G. gratefully acknowledges support from the Canada Council for the Arts  through its Killam Research Fellowship program. C.S. gratefully acknowledges a Goldhaber Distinguished Fellowship from Brookhaven Science Associates. Finally, we would like to thank Bj\"oern Schenke for helpful discussions.



%% References


\nocite{*}
\bibliographystyle{elsarticle-num}
\bibliography{jos}






% \end{thebibliography}

\end{document}